\documentclass[12pt]{article}

\def\ii{\'{\i }}
\usepackage{graphics}
\usepackage{epsfig}
\usepackage{fancyhdr}
\usepackage{amssymb}
\begin{document}
\renewcommand{\thefootnote}{\arabic{footnote}}
\title{Particle production azimuthal asymmetries in a clustering of color sources model}
\author{ I. Bautista$^{\ddag}$, L. Cunqueiro
\footnote{Laboratori Nazionali di Frascati, Via E.Fermi 40, 00044
Roma} $^{\ddag}$, J. Dias de Deus \footnote{CENTRA, Departamento
de F\ii sica, IST, Av. Rovisco Pais, 1049-001 Lisboa, Portugal}
 $^{\ddag}$ and C. Pajares \footnote{IGFAE and Departamento
de F\ii sica de Part\ii culas, Univ. of Santiago de Compostela,
15782, Santiago de Compostela, Spain}}
 \maketitle
\begin{abstract}

The collective interactions of many partons in the first stage of
the collisions is the usual accepted explanation of the sizable
elliptical flow. The clustering of color sources provides a
framework of partonic interactions. In this scheme, we
show a reasonable agreement with RHIC data for $p_{T}<1.5$ GeV/c
in both the dependence of $v_{2}$ on transverse momentum and in
the shape of the nuclear modified factor on the azimuthal angle
for different centralities.
We show the predictions at LHC energies for Pb-Pb. In the case of
proton-proton collisions a sizable $v_{2}$ is obtained at this
energy.
\end{abstract}
\section{Introduction}
A major breakthrough was the discovery by RHIC experiments of a
large elliptic flow $v_{2}$
\cite{Adcox:2004mh}\cite{Adams:2005dq}\cite{Manly:2005zy}\cite{Adcox:2002ms}
\cite{Adler:2004cj}\cite{Adler:2001nb}\cite{Adams:2003am}\cite{Back:2004mh}\cite{B.Alver}.
A non vanishing anisotropic flow exist only if the particles
measured in the final state depend not only on the physical
conditions realized locally at their production point, but if
particle production does also depend on the global event geometry.
In a relativistic local theory, this non local information can
only emerge as a collective effect, requiring interactions between
the relevant degrees of freedom, localized at different points of
the collision region. In this sense, anisotropic flow is a
particularly unambiguous and strong manifestation of collective
dynamics in heavy ion collisions \cite{Borghini:2007ub}. The large
elliptic flow $v_{2}$ can be qualitatively understood as follows:
In a collision at high energy, the spectators are fast enough to
free the way, leaving behind at mid- rapidity an almond shaped
azimuthally asymmetric region of QCD matter. This spatial
asymmetry implies unequal pressure gradients in the transverse
plane, with a larger density gradient in the reaction plane
(in-plane) perpendicular to it. As a consequence of the subsequent
multiple interactions between degrees of freedom, this spatial
asymmetry leads to an anisotropy in momentum space; the final
particle transverse momentum are more likely to be in-plane than
out-plane, hence $v_{2}>0$ as it was predicted
\cite{Ollitrault:1992bk}.

 The experimental results show that the elliptical flow first
increases with $p_{T}$ and then levels off around $p_{T} \thicksim
2$ GeV. Also, the $p_{T}$- dependent $v_{2}$ of identified hadrons
shows mass ordering at small $ p_{T}$ and displays a constituent-
quark counting rule. On the other hand $v_{2}$ normalized to the
eccentricity scales on the density of produced charged particles
to the overlapping area of the collision. To understand these
experimental results, many theoretical approaches have been used.
These include hydrodynamics models \cite{Huovinen:2001cy},
transport models \cite{Bravina:2004td}, and hybrid models which
combines both hydrodynamic and transport models
\cite{Teaney:2000cw}. Hydrodynamics models usually gives the
largest elliptical flow.

 In transport models, including the parton cascade \cite{Molnar:2001nk},
 it is obtained a
sizable $v_{2}$ only if it is used a very large parton cross
section. On the other hand, transport or hybrid models based on
string degrees of freedom in general gives a smaller elliptic flow
than the observed at RHIC \cite{Bleicher:2007cs}.

 The inclusion of interaction among strings via string fusion as
in the AMPT model\cite{Lin:2002cs}, in dual parton model
\cite{Bopp:2004xn} in the quark gluon string model
\cite{Bleibel:2006xx} or in Reggeon theory \cite{Boreskov:2009zz}
yield a large elliptic flow. In this paper, we would like to study
this point further, using percolation of strings. In this model,
is introduced the interaction between many degrees of
freedom. In fact  the overlapping of the strings forming color
clusters, can be seen as an interaction among the partons of the
projectile and target, from which the strings are stretched. The
recombination of the flavour and color of these partons give rise
to the flavours and color content of the resulting cluster. Due to
that, the fragmentation of the clusters is different of the
fragmentation of the initial strings, corresponding to a higher
color and also different flavour. This fact, gives rise to a
suppression of the multiplicities, a moderate increase of the
transverse momentum and an increase of the ratio baryon/meson with
centrality. Also it provide us with an state which has strongly
interacted in the first initial times. The percolation of strings
incorporates in a dynamical way two of the main
features of AA collisions, namely recombination and the formation
of a strong interacting partonic system. We will compare the
results obtained for the elliptic flow with experimental RHIC data
on AA collisions, including the dependence of the nuclear modified
factor on the azimuthal angles for different transverse momentum.
We will show our predictions for LHC energies, which gives a
slightly larger elliptic flow for all particles. A proton can be
considered as an extended object, although of smaller size than a
nucleus. In proton- proton collisions at LHC energies a high
number of strings are stretched between the partons of similar
effects to those seen at RHIC experiments could be expected.

 At first sight the difference between $p_{T}$
spectra observed in pp and AA collisions at RHIC energy would
suggest little flow, a conclusion in line with naive expectations.
However, it has been shown \cite{Chajecki:2008yi} that the effects
of energy and momentum conservation actually dominate the observed
systematic and that pp collisions may be much more similar to
heavy ion collisions than generally thought. In the framework of
percolation of strings we derived an approximate analytical
formula for the $p_{T}$ distributions, valid for AA and pp
collisions. Using this formula we compute the elliptic flow for pp
collisions, at LHC energy, obtaining a sizable size. This is not
unexpected because the string density at LHC energy in pp
collisions is close to the critical percolation density and
therefore there are many strings which overlap each other, giving
rise to high density partonic medium.

 The plan of the paper is as follows: In the next section, we
introduce the percolation of strings, computing the elliptic flow
in section 3. In section 4 we discuss our results comparing with
the RHIC data and show our predictions at LHC, for AA and  pp
collisions. Finally we present our conclusions.

\section{The string percolation model}
 In the string percolation model \cite{Armesto:1996kt}
 \cite{Pajares:2005kk} \cite{Mohring:1992wm}, multiparticle
production is described in terms of color strings stretched
between the partons of the projectile and the target. These
strings decay into new ones by $q\bar{q} $ or $ qq+
\bar{q}\bar{q}$ pair production and subsequently hadronize to
produce hadrons. Due to the confinement, the color of these
strings is confined to small area in transverse space $S_{1}=\pi
r_{0}^{2}$ with $r\backsimeq 0.2-0.3$ fm. These values comes from
lattice results on bilocal correlators \cite{Campostrini:1984xr}
and from the method of field correlators \cite{Dosch:1988ha}
\cite{DiGiacomo:2000va} and corresponds to the correlation length
of the QCD vacuum. With increasing energy and/or atomic number of
the colliding particles, the number of strings, $N_{s}$, grows and
they start to overlap forming clusters, very much like disks in
two dimensional percolation theory. At a certain critical density,
a macroscopical cluster appears, which marks the percolation phase
transition. This density corresponds to the value
$\rho_{c}=1.18-1.5$, (depending on the type of employed profile
functions of the nucleus, homogeneous or Wood- Saxon
\cite{DeDeus:2002qb}) where $\rho=N_{s}S_{1}/S_{A}$ and $S_{A}$
stands for the overlapping area of the colliding objects. The
percolation theory governs the geometrical pattern of the string
clustering. Its observable implications, however, require the
introduction of some dynamics in order to describe the behavior of
the cluster formed by several overlapping strings. We assume that
a cluster of n strings behaves as a single string with an
energy-momentum that corresponds to the sum of the energy-momenta
of the individual strings and with a higher color field,
corresponding to the vectorial sum of the color field of each
individual string. In this way \cite{Braun:2000hd}, we can compute
the mean multiplicity $\mu_{n}$ and the mean transverse momentum
squared $<p_{T}^{2}>_{n}$ of the particles produced by a cluster.
\begin{equation}
<\mu_{n}>=\sqrt{\frac{n S_{n}}{S_{1}}}<\mu_{1}> \mbox{  and  }
<p_{T n}^{2}>=\sqrt{\frac{n S_{1}}{S_{n}}}<p_{T 1}^{2}>,
\end{equation}
where $<\mu_{1}>$ and $<p_{T1}^{2}>$ are the mean multiplicity and
mean $p_{T}^{2}$ of particles produced by a single string. We take
$S_{1}$ constant equal to a disc of radius $r_{0}$. $S_{n}$
corresponds to the total area occupied by n discs, which of course
can be different for different configurations even if the clusters
have the same number of strings. Notice that if the strings are
just touching each other $S_{n}=nS_{1}$ and the strings act
independently to each other.On the contrary if they  fully overlap
$S_{n}=S_{1}$ it is obtained the largest suppression of the
multiplicities and the largest increase of the transverse
momentum. In the limit of high density, one obtains
\cite{Braun:2000hd}
\begin{equation}
<n\frac{S_{1}}{S_{n}}>=\frac{\rho}{1-e^{-\rho}}=\frac{1}{F(\rho)^{2}}
\end{equation}
and the equations (1) transform into the analytical ones.
\begin{equation}
<\mu>=N_{s}F(\rho)<\mu_{1}>, \mbox{   and  }
<p_{T}^{2}>=\frac{<p_{T 1}^{2}>}{F(\rho)}
\end{equation}
with $ F(\rho)=\sqrt{\frac{1-e^{-\rho}}{\rho}}$. In principle, the
equality (3) has been obtained in the high energy limit, what
means close or above, the critical density, where the strings
covers a large fraction of the total available area. However the
equations (3) give also the right behavior in the limit of low
density. in fact if $\rho$ tends to zero, $F(\rho)$ is one and
therefore there is not any change from the result of the
superposition of individual strings. In the mid rapidity region,
the number of strings is proportional to the number of collisions,
$N_{coll}\sim N_{A}^{\frac{4}{3}}$ being $N_{A}$ the number of
participants. Therefore $\rho\sim N_{A}^{\frac{2}{3}}$ and $\mu
\sim N_{A}$. In other words, the multiplicity per participant
becomes independent of $N_{A}$, i.e. saturates. Outside the mid
rapidity, $N_{s}$ is proportional to $N_{A}$ instead of
 $N_{A}^{\frac{4}{3}}$.
Therefore, there is an additional suppression factor
$N_{A}^{\frac{1}{3}}$ compared to central rapidity.

 If we are interested in a specific kind of particle i, we will use equations
(1) and (3) $\mu_{1i}$, $<p_{T1}^{2}>_{i}$, $<\mu_{n}>_{i}$ and
$<p_{Tn}^{2}>_{i}$ for the corresponding quantities.

In order to compute the multiplicities we must know, $N_{s}$ and
$\mu_{1}$ (for a fixed centrality, knowing $N_{s}$ we deduce the
density $\rho$ ). Up to RHIC energies, $N_{s}$ in the central
rapidity region is approximately twice the number of collisions,
$N_{coll}$. However $N_{s}$ is larger than $N_{coll}$ at RHIC and
LHC energies.

 Using equation (3) the multiplicities are computed
using a Monte-Carlo code based on the Quark- Gluon String model to
generate the strings \cite{Amelin:2001sk}. Each string is produced
at an identified impact parameter. From this, knowing the
transverse area of each string, we identify all the clusters
formed in each collision and subsequently compute for each of them
its multiplicity in units of $\mu_{1}$ the value of $\mu_{1}$ was
fixed by the comparison of our results with SPS data
 for Pb-Pb central collisions. Our results are in agrement with SPS
and RHIC multiplicity data \cite{Braun:2001us}. Using the first of
equations(1) we obtain very similar results
\cite{DiasdeDeus:2000gf}. Must of the reasonable string models
\cite{Amelin:2001sk} \cite{Capella:1992yb}\cite{Capella:1983ie}
\cite{Werner:1993uh}\cite{Andersson:1998tv} obtained similar
results for $N_{s}$. This fact gives us confidence is our values.
Notice that sometimes, even in experimental analysis, $N_{coll}$
is obtained from the optical approximation of the Glauber model
without taking into account the energy-momentum conservation. This
conservation reduces $N_{coll}$. In the evaluation of the
elliptical flow, we need the values of $dN/dy$ at central rapidity
for the different collisions and centralities, and values of
$\rho$. We take both from our previous results
\cite{Braun:2001us}\cite{DiasdeDeus:2000gf}.

Concerning the transverse momentum distribution, one needs the
distribution $g(x,p_{T})$ for each cluster, and the mean squared
transverse momentum distribution of the clusters $W(x)$, where x
is the inverse of the mean of the transverse momentum of each
cluster which is related to the cluster size by equation (3). For
$g(x,p_{T})$ we assume the Schwinger formula,
$g(x,p_{T}^{2})=exp(-p_{T}^{2}x)$ \cite{Pajares:2005kk}
\cite{DiasdeDeus:2003fg} used also as a first approximation for
the fragmentation of Lund string \cite{Andersson:1998tv}. For the
weight function, we assume the gamma distribution
 $W(x)=\frac{\gamma(\gamma x)^{k-1}}{\gamma(k)}e^{-kx)}$,\cite{DiasdeDeus:2003ei}
 \cite{Cunqueiro:2007fn} with $\gamma=\frac{k}{<x>}$ and
$\frac{1}{k}=\frac{(<x^{2}>-<x>^{2})}{<x>^{2}}$ is proportional to
the width of the distribution, depending on the density of strings
$\eta$. Therefore the
 transverse momentum distribution
$f(p_{T},y)$ of the particle i is
\begin{center}
$f(p_{T},y)=\frac{dN}{dp_{T}^{2}dy}=\int_{0}^{\infty}dxW(x)g(p_{T},x)$
\end{center}
\begin{equation}=\frac{dN}{dy}\frac{k-1}{k}\frac{1}{<p_{T}^{2}>_{1i}}F(\rho)\frac{1}
{(1+F(\rho)p_{T}^{2}/k<p_{T}^{2}>_{1i})^{k}}
\end{equation}
The equation (4) is valid for all types of collisions and high
densities. The equations (4) can been seen as superposition of
color sources, clusters, where $\frac{1}{k}$ fixes the transverse
momentum fluctuations. At small density of strings, there will not
be overlapping of strings, the strings are isolated and
$k\rightarrow\infty$. When $\rho$ increases, there are some
overlapping of strings, forming clusters and therefore k
decreases. The minimum of k will be reached when the fluctuations
reaches its maximum. Above this point,
 increasing $\rho$, these fluctuations
decrease and k increases. In the limit of $\rho\rightarrow\infty$,
there is only one cluster formed by all the produced strings.
Again, there are no fluctuations and k tends to infinity.

The quantitative dependence of k on $\rho$ was obtained in
 \cite{DiasdeDeus:2003ei} from the comparison of equation (4) with
 RHIC AA data at different centralities. We use here such a
dependence. The peripheral Au-Au collisions at $\sqrt{s}=200$ GeV,
corresponds to the $\rho$ values slightly above the minimum of k
\cite{DiasdeDeus:2003ei}. In the range of the used $\rho$ values,
k is a smooth function on $\rho$. We will show that for $p_{T}<
1.5 $ GeV/c our $v_{2}$ does not depend on k. The formula (4) is
able to describe the transverse momentum distribution for mesons
in AA collisions at all centralities and energies up to moderate
$p_{T}$, $p_{T}<4-5$ GeV/c. In particular they describe rightly
the Cronin effect and the suppression of moderate $p_{T}$
particles, as it was shown in reference \cite{DiasdeDeus:2003ei}.

We do not claim to describe the data at high $p_{T}$. It is well
known that jet quenching is the mechanism responsible for the high
$p_{T}$ suppression. This phenomena is not included in
  our formula (4) which was obtained
assuming a single exponential for the decay of a cluster without a
power like tail. Our formula can be seen as a way to join smoothly
the low and moderate $p_{T}$ region with the high $p_{T}$ region.
 Indeed, the high $p_{T}$
suppression implies by continuity a suppression of the highest
$p_{T}$ values of the
 intermediate region which are
described by formula (3).

 The universal formula (4) must be
considered as an analytical approximation to a process that
consists on the fragmentation of clusters of strings and their
eventual decay via successive cluster breaking via Schwinger
mechanism \cite{Amelin:1994mc}. The formula (4) is not able to
explain the differences between antibaryons (baryons) and mesons.
In fact, the only parameter that is different for them in (4) is
the mean of the squared of transverse momentum of pions and
protons produced by a single string $<p_{T}^{2}>_{1\pi}$ and
$<p_{T}^{2}>_{1p}$, respectively. This only causes a shift on the
maximum of $R_{AA}$ but keeps the same height at the maximum
contrary to what one observed. However, in the fragmentation of a
cluster formed by several strings, the enhancement of production
of antibaryons (baryons) over mesons is not only due to a mass
effect corresponding to a higher tension due to a higher density
of the cluster (the factor $F(\rho)$ in front of $p_{T}^{2}$ in
(4)). In fact, the color and flavour properties of a cluster
follow from the corresponding properties of their individual
strings. A cluster composed by several quark-antiquarks
$(q-\bar{q})$ strings behaves like an string with a color Q and
flavour composed of the flavour of the individual strings, as a
result, we obtain clusters with higher color and different flavour
ends. For the fragmentation of a cluster we consider the creation
of a pair of parton complex as $Q \bar{Q}$ \cite{Amelin:1994mc},
after the decay, the two new $Q \bar{Q}$ strings are treated in
the same manner and therefore decay into more $Q\bar{Q}$ strings,
until they come to objects with masses comparable to hadron
masses, which are identified with observable hadrons by combining
with them the produced flavour with statistical weights. In this
way, the production of antibaryons (baryons) is enhanced with the
number of strings of the cluster. The additional antiquarks
(quarks) required to form an antibaryon (baryon) are provided by
the antiquarks (quarks) of the overlapping strings that form the
cluster \cite{Amelin:1994mc}\cite{Amelin:1993cs}. Therefore
recombination or coalescence mechanism \cite{Hwa:2002tu}
\cite{Fries:2003vb} are incorporated in a dynamical way. In order
to take this into account keeping our close formula we will do the
following approximation. For antibaryons (baryons) we consider
instead of the equation (3),
\begin{equation}
\mu_{\bar{B}}=N_{s}^{1+\alpha}F({\rho_{\bar{B}}})\mu_{1\bar{B}}
\end{equation}
fitting the parameter $\alpha$ to reproduce the experimental
dependence of the $p_{T}$ integrated $\bar{p}$ spectra with
centrality \cite{Cunqueiro:2007fn}. The obtained value is
$\alpha=0.09$. this means that the density $\rho$ must be replaced
by $\rho_{\bar{B}}=N_{s}^{\alpha}\rho$ The antibaryons (baryons)
probe a higher density than mesons for the same energy and type of
collision. The equation (4)  with the described modification for
baryons provide us the input for the evaluation of the elliptic
flow.

\section{Elliptic Flow.}

In saturation initial state models as the color Glass Condensate
 \cite{McLerran:1993ni} \cite{Kovchegov:2002nf} or the Percolation
String Model \cite{Armesto:1996kt} \cite{Pajares:2005kk} color
sources radially emit particles. The azimuthal asymmetry has, in
this case, the origin in the deviation from the circle of the
impact parameter projected interacting overlap region in non
central, $b>0$, collisions. The situation is schematically
represented in Fig. 1. If one measures production within fixed
azimuthal angles, one has to know the source density as a function
of $R_{\varphi}$. If no azimuthal determinations are made one can
adopt a circle of radius R such that
\begin{equation}
\frac{\pi
R^{2}}{4}=\frac{1}{2}\int_{0}^{\frac{\pi}{2}}R_{\varphi}^{2}d{\varphi}
\end{equation} or \begin{equation}
R^{2}=<R_{\varphi}^{2}>
\end{equation}
Note that, as in the experiment, we are using for the range of
variation of $\varphi$, $0<\varphi<\frac{\pi}{2}$. As the number
of color sources is the same in the interior of the circle and in
the ellipsoid and as the directional density is related to the
average density by the formula
\begin{equation}
\rho_{\varphi}=\rho(\frac{R}{R_{\varphi}})^{2}
\end{equation}
one expects higher $<p_{T}^{2}>$ and larger particle numbers for
production in the reaction plane, $\varphi=0$. In order to make
this point clear let us consider a model, simpler than (4), which
is the Schwinger model with percolation \cite{DiasdeDeus:2003fg}:
\begin{equation}
\frac{2}{\pi}\frac{dN}{dp_{T}^{2}}\sim
e^{-F(\rho)\frac{p_{T}^{2}}{<p_{T}^{2}>_{1}}}
\end{equation}
for the azimuthal integrated cross section and
\begin{equation}
\frac{dN}{dp_{T}^{2}d\varphi} \sim
e^{-F(\rho_{\varphi})\frac{p_{T}^{2}}{<p_{T}^{2}>_{1}}}
\end{equation}
for a direction $\varphi$. From (9) and (10) one sees that the
ratio of the $p_{T}$ integrated quantities is
\begin{equation}
%\frac{N(p_{T}^{2})\varphi}{\frac{2}{\pi}N(p_{T}^{2})}=
\frac{(10)}{(9)}=
\frac{<p_{T}^{2}>_{\varphi}}{<p_{T}^{2}>}=\frac{F(\rho)}{F(\rho_{\varphi})}
\end{equation}
which for small $\varphi$, as $F(\rho)$ is decreasing function of
$\rho$, is larger than 1. The same result works for Eq. (4). The
relevant directional factor is the factor $F(\rho)p_{T}^{2}$ or
$F(\rho_{\varphi})p_{T}^{2}$. Let us define, for a given
centrality, number of participants or $\rho$,
\begin{equation}
{dn\over dy\ dp_{T}^2}\ \vert_{y=0} \equiv f\left(F(\rho),p_{T
}^2\right)
\end{equation}
and
\begin{equation}
{dn\over dy\ dp_{T}^{2} d\varphi} |_{y=0} \equiv f \left(
F(\rho_{\varphi}),p_{T}^2 \right)
\end{equation}
We will treat the azimuthal distribution as perturbation around
the average, isotropic distribution. Such expansion, as $\rho$ (
is proportional) to $\frac{1}{R^{2}}$ and $\rho_{\varphi}$ to
$\frac{1}{R_{\varphi}^{2}}$, can be written as:

\begin{figure}
\begin{center}
      \resizebox{60mm}{!}{\includegraphics{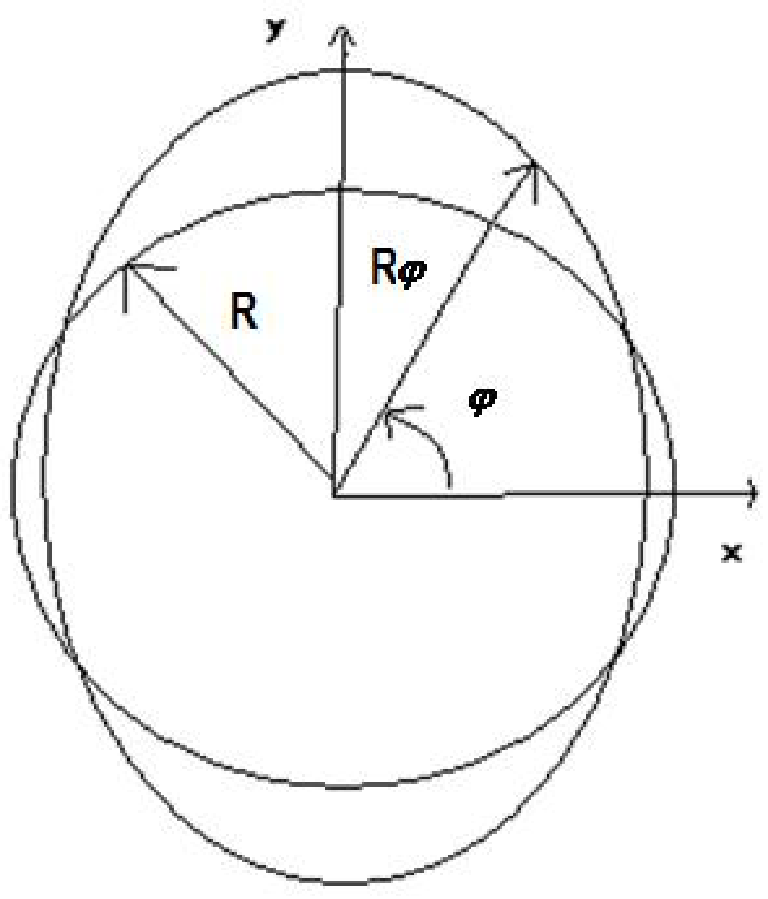}}
    \caption{}
    \label{test4}
\end{center}
\end{figure}

\begin{equation}
f\left( F(\rho_{\varphi}),p_{T}^2 \right) \simeq {2\over \pi}
 f\left( F(\rho),p_{T}^2 \right) \left[ 1 + {\partial \ln f\left( p_{T}^2 , R^2\right)
  \over \partial R^2}\left( R_{\varphi}^2-R^2 \right) \right]
\end{equation}
Note that (4), because of (6), satisfies the normalization
condition,
\begin{equation}
\int_0^{\pi/2} f\left(F(\rho_{\varphi}), p_{T}^2 \right) d\varphi
= f\left(F(\rho), p_{T}^2 \right)
\end{equation}
Additional terms in the expansion will, in general, violate the
normalization condition (15). As a decrease in $R^{2}$ corresponds
to an increase in source density and an increase in particle
production one expects, in (16)
\begin{equation}
\frac{df(F(\rho),p_{T}^{2})}{dR^{2}}<0
\end{equation}
with the consequence, again see (16), that production is maximal
for $\varphi\sim 0$. Note that the second term in the right hand
side of (16), changes sign at $R_{\varphi}=R^{2}$.

We shall next write expressions for relevant measurable quantities
as modification factors, $R_{AA}$ and $v_{2}$. With the usual
definition for $R_{AA}(p_{T}^{2})$, we define,
\begin{center} $R_{AA} (p_{T}^2 ,
\Delta \varphi) \equiv \Delta \varphi
{f_{AA}(F(\rho_{\varphi})p_{T}^2)\over \langle N_{coll}\rangle
f_{pp} (F(\rho)p_{T}^2)}$
\end{center}
\begin{equation}
= \Delta \varphi R_{AA} (p_{T}^2) \left[ 1+ {\partial \ln
f_{AA}(F(\rho),p_{T}^2) \over\partial R^2}\left(R_{\varphi}^2 -
R^2 \right)\right]
\end{equation}
and
\begin{equation}
v_2 \left( p_{T}^2 \right) = {2\over \pi} \int_0^{\pi \over 2}
d\varphi \ cos (2 \varphi) \left[ 1+ {\partial \ln f_{AA}
(F(\rho),p_{T}^2 )\over \partial R^2}\left( R_{\varphi}^2 - R^2
\right)\right]
\end{equation}
Note that (16), using (4), becomes
\begin{equation}
f\left( p_{T}^2 , R_{\varphi}^2 \right) = f\left( p_{T}^2 , R^2
\right)\left[ 1+ {1\over R^2 F(\rho )} \left( e^{-\rho} -
F(\rho)^2 \right) {p_{T}^2 \over 2\overline{p_{1}^2}} {1\over 1+
{F(\rho) p_{T}^2\over k \overline{p_{1}^2}}} \left(
R_{\varphi}^2-R^2 \right) \right]
\end{equation}
From the equations (19), (20), (21) we can compute our main
observables to compare with the experimental data. The
approximation done is valid when the second term in (21) is small,
what means not hight $p_{T}$. Notice that due to the location in
impact parameter space of the strings that form clusters there is
an initial spatial azimuthal asymmetry which is translated into
transverse momentum due to interactions of strings included in the
formula for the $p_{T}$ distribution via the factor $F(\rho)$. The
strength of the interaction depends locally on the string density
and therefore is the origin of the azimuthal asymmetry.

\section{Results and Discussion}

In order to compute the elliptic flow, we need the values of
$N_{s}$, $\rho$, $k(\rho)$ and $<p_{T}^{2}>_{1\pi}$,
$<p_{T}^{2}>_{1k}$,  $<p_{T}^{2}>_{1\bar{p}}$. As we mentioned
before, for each type of collision, energy and centrality, we
obtain the number of strings using a Monte-Carlo code based on the
Quark Gluon string model \cite{Amelin:2001sk}, hence we compute
$\rho$ and $k(\rho)$. The  dependence of k on $\rho$ is very
smooth, our results for $p_{T}<1.0$ GeV/c are independent  of k
according to formula (18) and (19). In table 1, we show the value
of $N_{s}$, $\rho$, $k(\rho)$ for different types of collisions,
centralities and energies. We include also the values of the
charged multiplicity at central rapidity $\frac{dN}{dy}$ which we
will use below. They are taken from previous work done in the
framework of percolation \cite{Braun:2001us}
\cite{DiasdeDeus:2000gf}
\cite{DiasdeDeus:2003fg}\cite{p.brogueira}.
 We use the values
$<p_{T}^{2}>_{1\pi}=0.06(GeV/c)^{2}$,
$<p_{T}^{2}>_{1k}=0.14(GeV/c)^{2}$,
$<p_{T}^{2}>_{1\bar{p}}=0.30(GeV/c)^{2}$, obtained previously from
the comparison of formula (4) with the experimental data of the
dependence of the mean transverse momentum of the different
particles on the multiplicity \cite{DiasdeDeus:2003fg}
\cite{DiasdeDeus:2003ei}.

Once these parameters are fixed then the elliptic flow for a
certain type of collision, centrality and energy is known because
it only depends on $\rho$. Notice, that we could consider
$<p_{T}^{2}>_{\pi}$, $<p_{T}^{2}>_{1k}$, $<p_{T}^{2}>_{1\bar{p}}$
as free parameters doing a fit to the experimental data. However,
our aim is not to obtain a perfect agreement with data. We are
aware of some simplifications and limitation of our framework,
therefore we just check whether the general trend of the data is
reproduced in our model.

In fig. 2 we show our results for minimum bias Au-Au collisions
for pions, kaons and antiprotons compared with the experimental
data \cite{Adler:2003kt}. A reasonable agreement is obtained for
$p_{T}\leq 1$ GeV/c, reproducing the mass ordering. In fact, at
low $p_{T}$, the elliptic flow is proportional to $p_{T}^{2}$ and
the proportionality coefficient is
$\frac{1}{<{p}_{T1}^{2}>_{\pi}}$. As $<p_{T1}^{2}>_{\pi}<
<p_{T1}^{2}>_{k}< <p_{T1}^{2}>_{\bar{p}}$ $v_{2}$ is larger for
pions than for kaons and antiprotons. Our evaluation was done
retaining only the first term of the expansion in powers of
$R_{\phi}-R$. For $p_{T}<1.5$ GeV/c, this term is small, however
for larger $p_{T}$ we should include additional terms.

\begin{figure}
\begin{center}
      \resizebox{120mm}{!}{\includegraphics{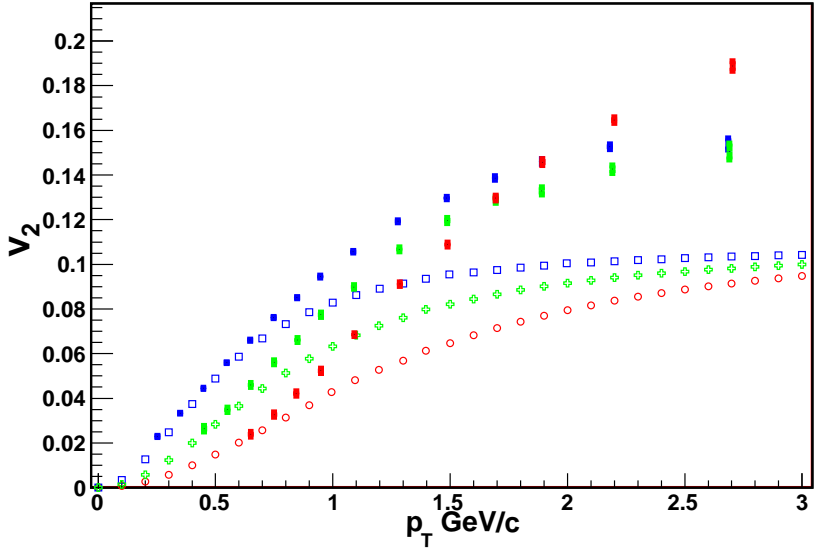}}
    \caption{Elliptic flow for different hadron species in Au+Au collisions, color lines in red, blue and green show our predictions, for antiprotons, pions and kaons respectively, dashed lines experimental data from RHIC \cite{Adare:2006ti}. }
    \label{test4}
\end{center}
\end{figure}

In fig. 3 we compare our results on the azimuthal angle dependence
of the nuclear modified factor $R_{AA}(p_{T}^{2},\Delta\varphi)$
at different centralities with the PHENIX experimental data
\cite{:2009iv}. The model reproduces the shape for all
centralities although the results are below the experimental data
for low angles.

\begin{figure}
\begin{center}
      \resizebox{120mm}{!}{\includegraphics{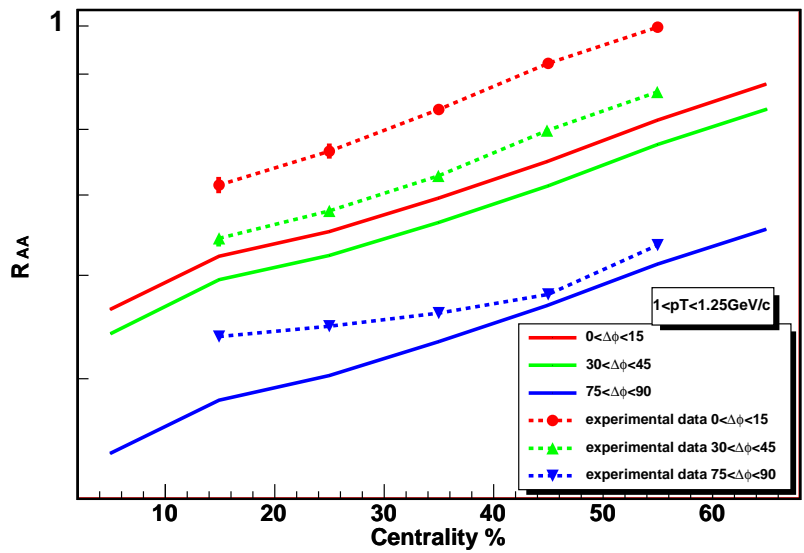}}
     \caption{Angular suppression for pions at different centrality, $(10-20\%, 20-30\%, 30-40\%, 50-60\%)$ results compared to PHENIX data.}
     \label{test4}
\end{center}
\end{figure}

In fig. 4 we plot $v2/\epsilon_{part}$ being
$\epsilon=\frac{\sqrt{(\sigma^{2}_{y}-\sigma^{2}_{x})^{2}+4(\sigma_{xy})^{2}}}{\sigma_{x}^{2}+\sigma_{y}^{2}}$
the participant eccentricity as a function of the number of
participants $N_{A}$ together the PHOBOS
\cite{Alver:2006wh}\cite{Adler:2007cb} experimental data in Cu-Cu
and Au-Au at $\sqrt{s}=200$ GeV. It is observed good agreement.
%We compare our results with Cu-Cu RHIC experimental data\cite{Alver:2006wh} at different centralities.

\begin{figure}
\begin{center}
      \resizebox{120mm}{!}{\includegraphics{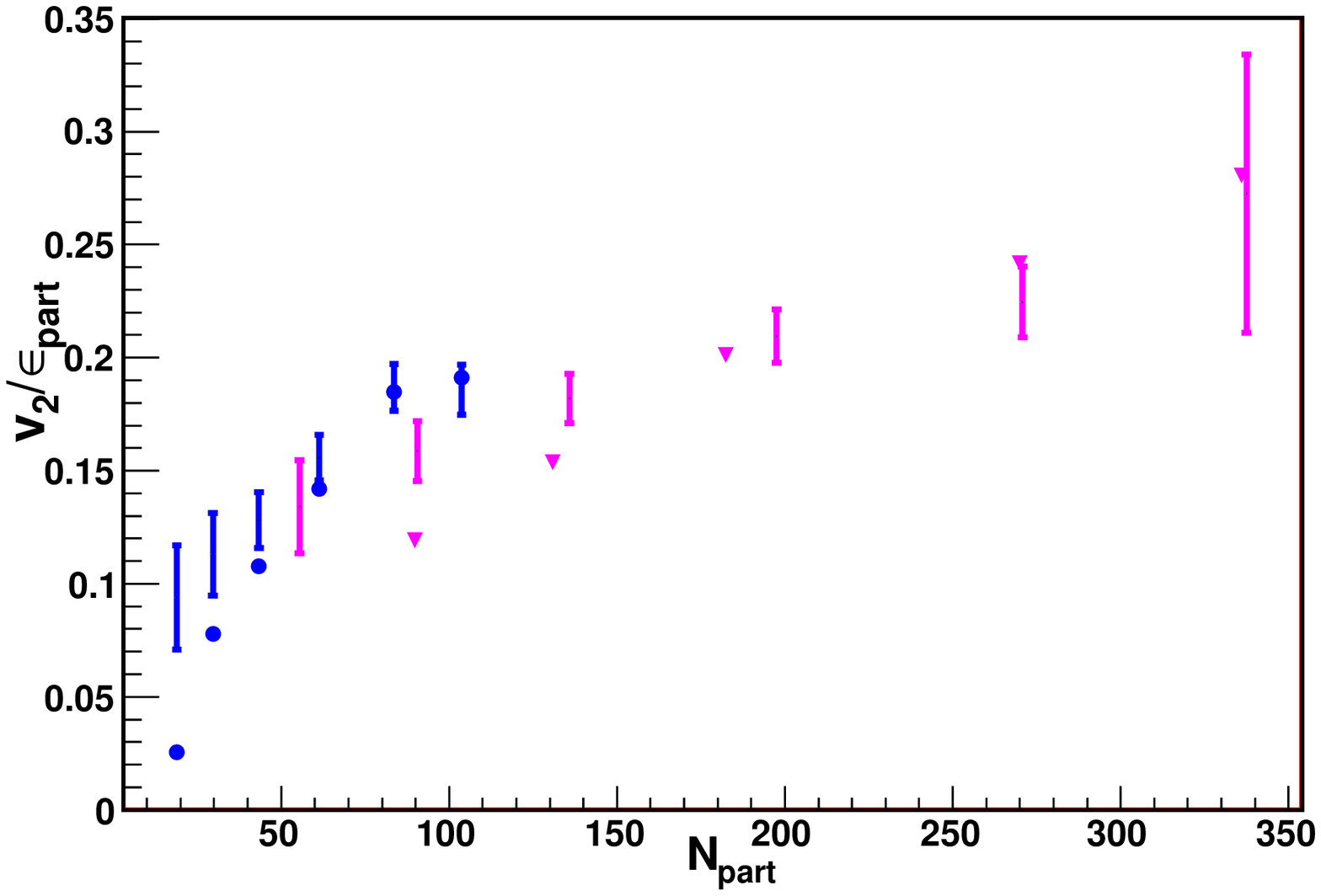}}
    \caption{Elliptic flow $v2/\epsilon_{part}$ as a function of the number of participants. Pink triangles and round points are the model results for Au-Au and Cu- Cu collisions at $\sqrt{s}=200$ GeV.}
    \label{test4}
\end{center}
\end{figure}
 In fig. 5 we plot $v_{2}/\epsilon$, being the
participant eccentricity as a function of the ratio
$\frac{dN_{ch}}{dy}\frac{1}{S}$ where S is the surface of the
collision for a given centrality. Together our results we show the
experimental data at AGS, SPS and RHIC energies for different type
of collisions \cite{Voloshin:2007af}. A good agreement is
observed. The scaling, experimentally observed, is also expected
in our model. In fact, from equation (2) we obtain.
\begin{figure}
\begin{center}
      \resizebox{120mm}{!}{\includegraphics{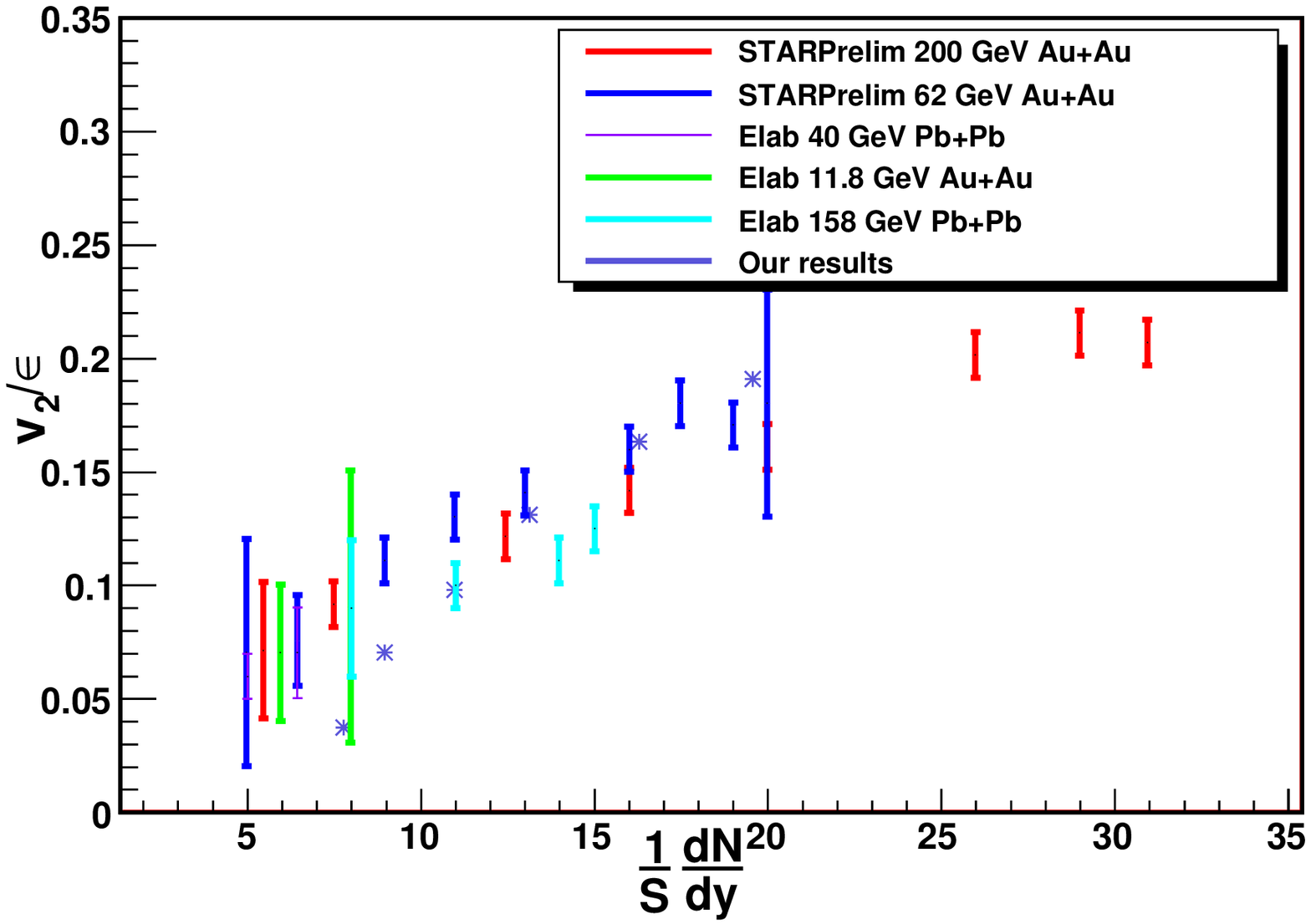}}
    \caption{Scaling predictions for Au+Au at $\sqrt{s}=200$ GeV (stars).}
    \label{test4}
\end{center}
\end{figure}

\begin{equation}
\frac{1}{S}\frac{dN_{ch}}{dy}=\rho
F(\rho)\frac{\mu_{1}}{r_{0}^{2}}
\end{equation}
which is proportional to $\sqrt{\rho}$ at high densities. On the
other hand, from equations (22) and (20) we have
\begin{equation}
v_{2}(p_{T})=A(\frac{2}{\pi}\int_{0}^{\frac{\pi}{2}}d\phi
cos(2\phi)(\frac{R_{\phi}^{2}}{R^{2}}-1))
\end{equation}
with
\begin{equation}
A=\frac{e^{-\rho}-F(-\rho)^{2}}{F(\rho)}
\frac{p_{T}^{2}}{2\bar{p}_{T1}^{2}}\frac{1}{1+\frac{F(\rho)p_{T}^{2}}
{2\bar{p}_{T1}^{2}}}
\end{equation}

As the bracket of the equation(24) is related to the centrality
and A depends only on $\rho$, i.e. depends only on
$\frac{1}{S}\frac{dN_{ch}}{dy}$. The equation (21) gives
approximately the scaling law observed.

In fig. 6 we plot the function $-(e^{-\rho}-F(\rho)^{2})/F(\rho)$
as a function of $\rho$. We observe that increases up to
$\rho\approx 2.7$ where it starts to decrease slowly. Therefore we
predict this behavior for the elliptic flow. As the value
$\rho=2.5$ only is reached for central Au-Au collisions at RHIC or
semicentral at LHC energies, only for this type of collisions will
be sightly smaller at LHC energies. As the antibaryons(baryons)
probe a higher string density than the mesons the elliptic flow
for them will decrease for a less degree of centrality.
\begin{figure}
\begin{center}
      \resizebox{120mm}{!}{\includegraphics{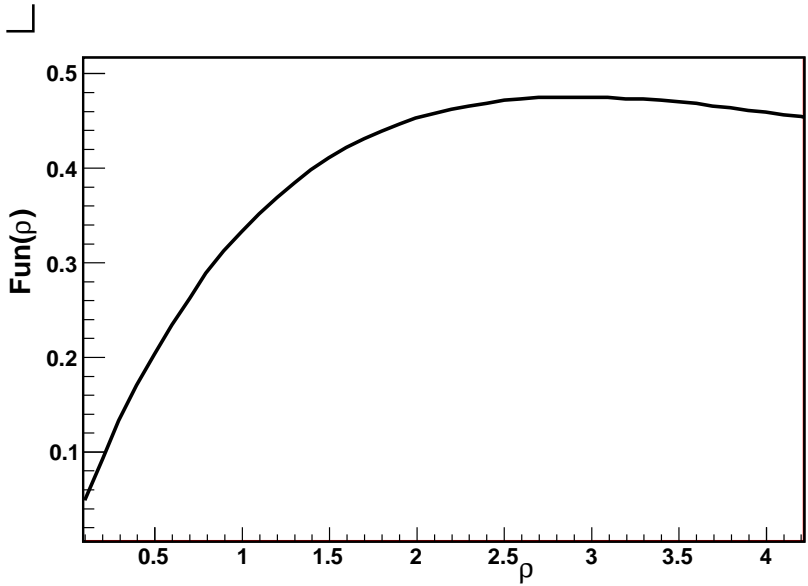}}
    \caption{Dependence on the string density of the function described in the text}
    \label{test4}
\end{center}
\end{figure}

In fig. 7 we plot $v_{2}/nq$ as a function of the kinetic energy
over the constituent quarks of the observed particles. The
recombination models \cite{Hwa:2002tu}\cite{Fries:2003vb}
predicted an scaling law for intermediate transverse momentum,
independently of mesons or baryons, all particles should lay in
the same universal curve. We show our results for pions, kaons and
antiprotons together with the experimental data
\cite{Adare:2006ti}. It is seen that our model satisfies the
scaling law as it was expected. In fact, in the percolation of
strings, in addition to the mass effect due to a higher tension of
the cluster which gives rise to enhancement of heavier particles,
there is a flavouring effect related to the number of strings of
the cluster, each one with flavour, which gives rise to a higher
probability to combine them, producing more baryons
%\cite{Amelin:1994mc}\cite{Brogueira:2009nj}
\cite{Cunqueiro:2007fn} \cite{Amelin:1994mc}. In order to obtain
close formula we incorporate this, only in approximate way as was
explained before, however our results for pions, kaons and
antiprotons are laying approximately in the same curve as the
experimental data, for KeT/$nq<0.2$ GeV, corresponding to low
$p_{T}$. For higher $p_{T}$ the result is below the experimental
data.

\begin{figure}
\begin{center}
     \resizebox{130mm}{!}{\includegraphics{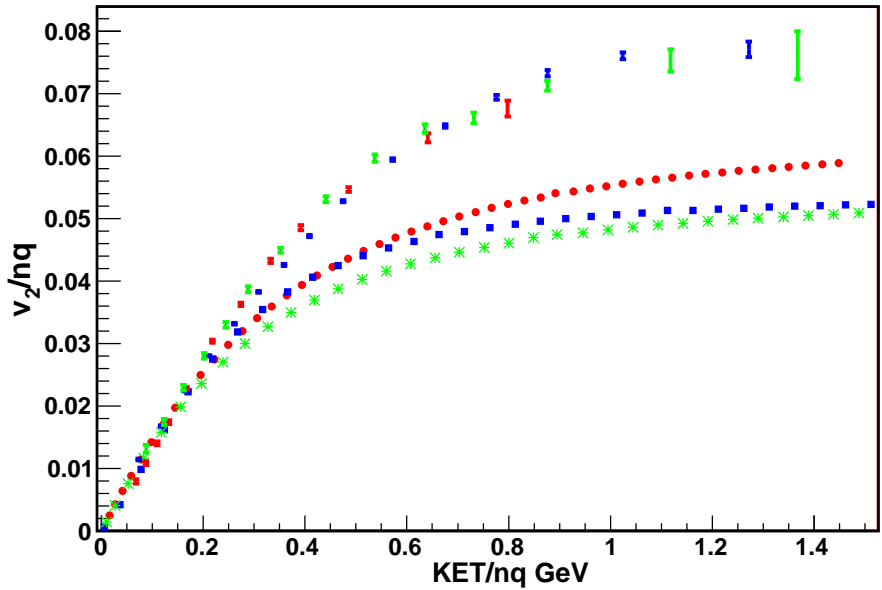}}
    \caption{The red, green and blue solid lines stand for the model results for antiprotons, kaons and pions respectively compared with the experimental data (errors are smaller than symbols used).}
    \label{test4}
\end{center}
\end{figure}

In figs. 8, 9, 10 we show our results at $\sqrt{s}=5.5$ TeV
compared with $\sqrt{s}=200$ GeV for different hadrons. For the
three particles we obtain a moderate increase of elliptic flow.
This is a consequence that the string density at both energies for
minimum bias Au-Au collisions is below $2.7$. Above this density,
the elliptic flow decreases what happened for Pb-Pb central
collisions at LHC.
\begin{figure}
  \begin{center}
      \resizebox{120mm}{!}{\includegraphics{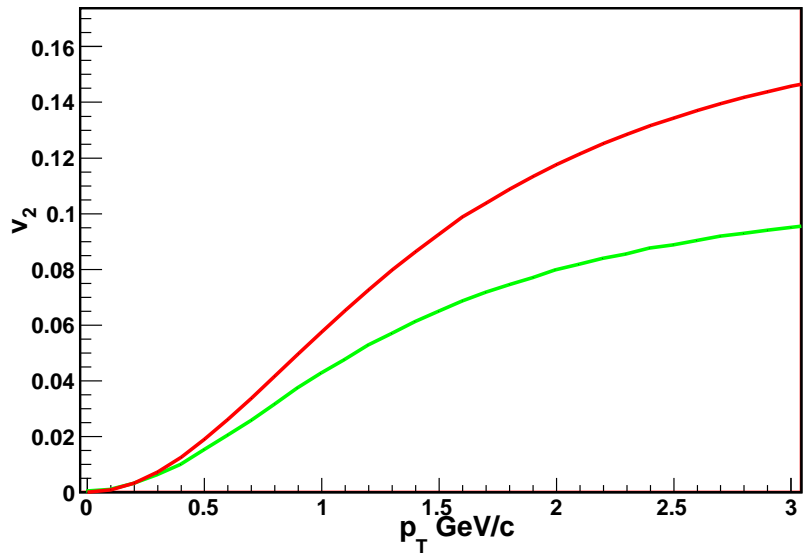}}
    \caption{Expected elliptic flow for antiprotons by our model, green lines RHIC and red lines LHC energies.}
    \label{test4}
  \end{center}
\end{figure}
\begin{figure}
  \begin{center}
      \resizebox{120mm}{!}{\includegraphics{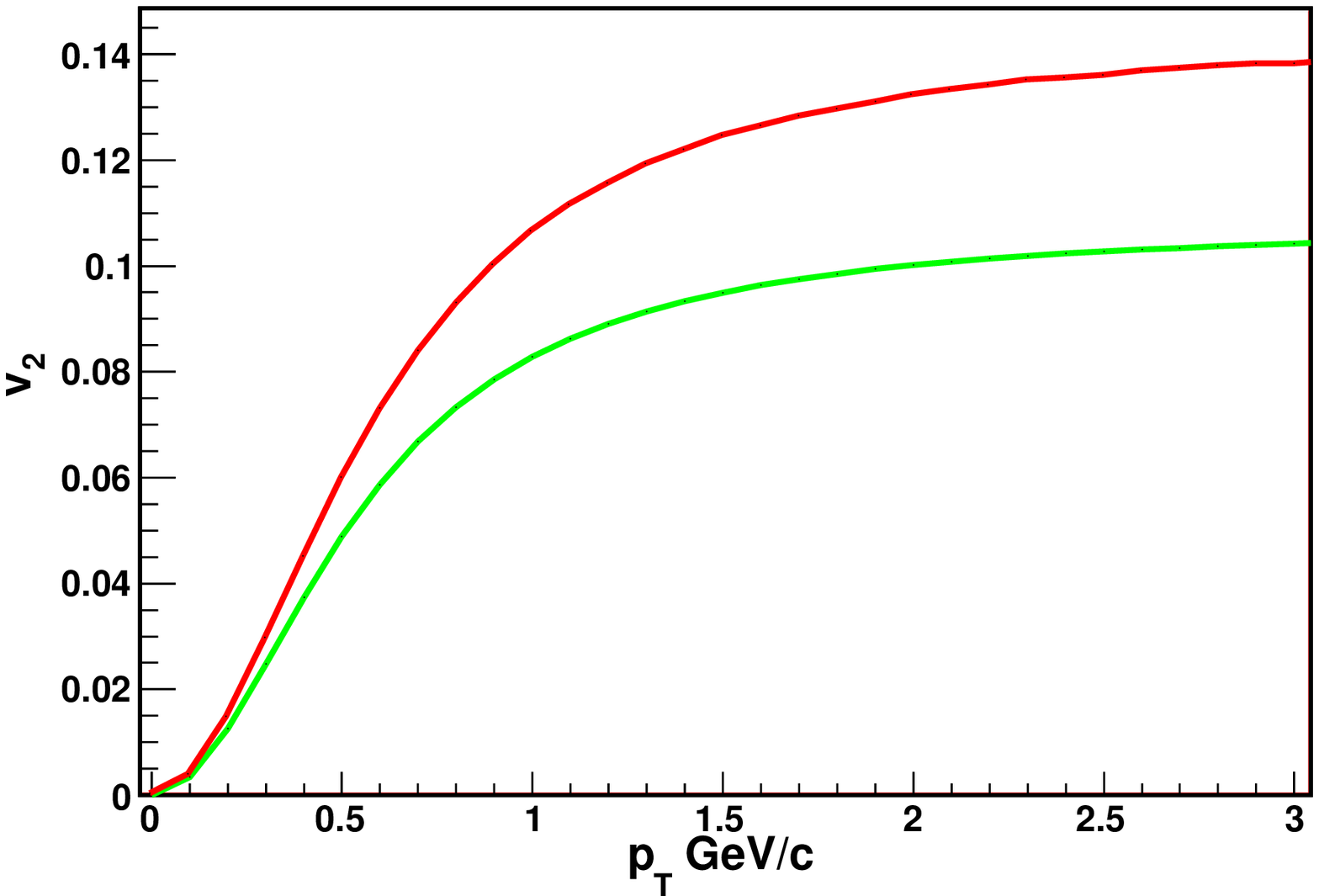}}
    \caption{Expected elliptic flow pions by our model, green lines RHIC and red lines LHC energies.}
    \label{test4}
  \end{center}
\end{figure}
\begin{figure}
  \begin{center}
      \resizebox{120mm}{!}{\includegraphics{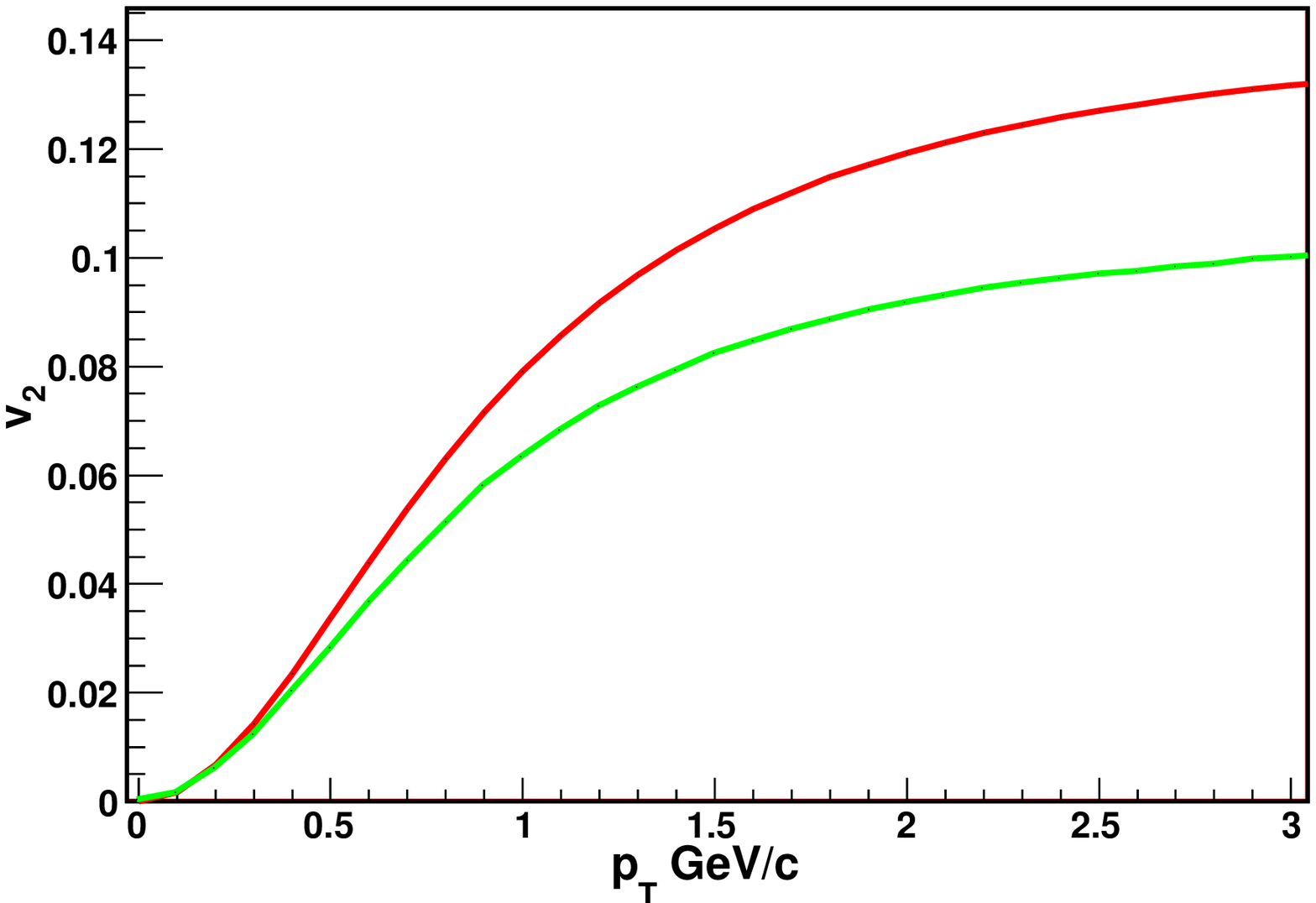}}
    \caption{Expected elliptic flow kaons by our model, green lines RHIC and red lines LHC energies.}
    \label{test4}
  \end{center}
\end{figure}
\begin{figure}
\begin{center}
      \resizebox{120mm}{!}{\includegraphics{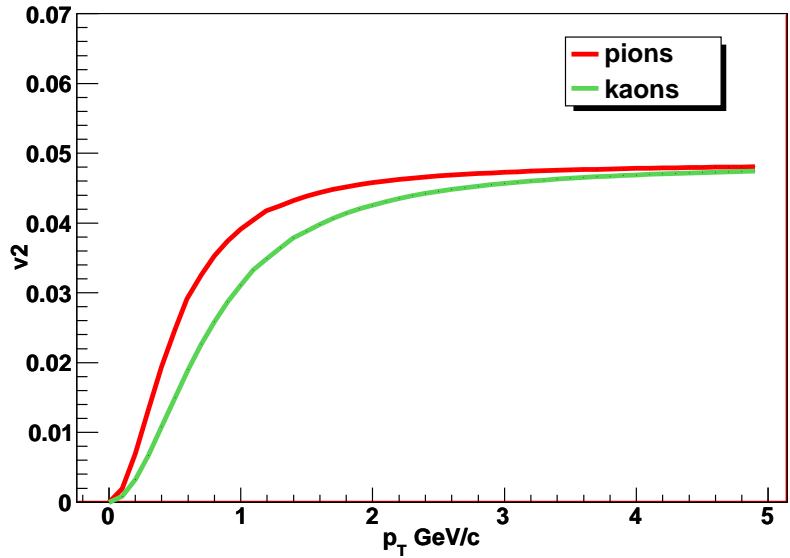}}
    \caption{pp predictions at $\sqrt{s}=14$ TeV.}
    \label{test4}
\end{center}
\end{figure}

In fig. 11 we show our prediction for $v_{2}$ for pp collisions at
$\sqrt{s}=14TeV$ for pions and kaons. A sizable elliptic flow is
obtained, this result is not surprising, given the high density
partonic structure obtained in pp at LHC which is similar, see
table 1, to Cu-Cu collisions The high density partonic structure
of pp at LHC can give rise to other effects observed in AA
collisions at RHIC as the high $p_{T}$ particles and the existence
of large rapidity long range correlation even at very large
intervals of rapidity
\cite{Cunqueiro:2008uu}\cite{Brogueira:2009nj}\cite{Armesto:2006bv}.

\section{Conclusions.}

In the framework of the clustering of color sources, the elliptic
flow, $v_{2}$, and the dependence of the nuclear modified factor
on the azimuthal angle have been evaluated.

The comparison with RHIC experimental data for $p_{T}<1.5$ GeV/c shows a reasonable
agreement. In particular, the observed mass ordering is
reproduced. In the model, is obtained an analytical expression for
the observed scaling of the elliptic flow normalized to the
participant eccentricity on the ratio $\frac{1}{S}\frac{dN}{dy}$
The analytical expression indicates that above a determined
density $\rho=2.7$ elliptical flow decreases slowly. This density
is reached in Au-Au central collisions both at RHIC and LHC
energies, therefore we predict slightly decrease in this case. For
Au-Au minimum bias the density is lower than 2.7 and we predict a
slightly increase in that case.

We have computed the dependence of the nuclear modify factor on
the azimuthal angle comparing our results with experimental data
for different centralities. We reproduce the general trend of the
data, although we are $15\%$ below for small angles.

 We predict sizable elliptic flow in proton proton collisions, consequence of
the high partonic density reached at LHC energy. The main
ingredient of the model is the interaction among strings or
equivalently among the partons of projectile and target located at
the end of each string. In this way, we have a high density
partonic interacting medium. The reasonable agreement obtained in
our model with the experimental data for $p_{T}<1.5$ GeV/c, confirms the common belief
that elliptic flow reveals a strongly interacting medium created
in the first stage of the collision. Our predictions will be
tested with the incoming LHC experiments.
\begin{center}
\begin{tabular}{|c|c|c|c|c|}
  \hline
  \  & $N_{s} $& $\rho$ & $k(\rho)$ & $\frac{dN}{dy}$\\
  \hline
  pp (m.b) & 8 & 0.4 & 3.6 & 2.5 \\
  \hline
  - & 14 & 0.7 & 3.4 & 5.5 \\
  \hline
  Cu-Cu (m.b) & 100 & 0.4 & 3.6 & 50 \\
  \hline
  - & 170 & 0.7 & 3.4 & 120 \\
  \hline
  Au-Au (m.b)  & 390 & 1.0 & 3.6 & 180 \\
 \hline
  - & 550 & 1.6 & 3.6 & 400 \\
  \hline
  Au- Au $(0-10)\%$ & 1600 & 2.7 & 4.0 & 650 \\
  \hline
  - & 2600 & 4.8 & 4.1 & 1500 \\
  \hline
\end{tabular}
\end{center}
\begin{center}
Table 1: Values of $N_{s}$, $\rho$, $k(\rho)$ and $\frac{dN}{dy}$
for pp, Cu-Cu, Au-Au minimum bias (m.b) and $0-10\%$ central Au-Au
collisions. For each collision the above numbers correspond to
$\sqrt{s}=200$ GeV and the below ones to $\sqrt{s}=5.5$ TeV.
\end{center}
\textbf{\Large Acknowledgements}
% \begin{Acknowledgement}

We thank Dr. N. Armesto and Dr. Carla Vale for discussions.

J. D. D. thanks the support of the FCT/Portugal project
PPCDT/FIS/575682004.

I. B, L. C. and C. P. were supported by the project FPA2008-01177
of MICINN, the Spanish Consolider –Ingenio 2010 program CPAN and
Conselleria Educacion Xunta de Galicia.

% \end{acknowledgements}

\end{document}